# Electric field enhancement of the superconducting spin-valve effect via strain-transfer across a ferromagnetic/ferroelectric interface

Tomohiro Kikuta[1], Sachio Komori[1]*, Keiichiro Imura[1,2], Tomoyasu Taniyama[1]*

[1]Department of Physics, Nagoya University, Furo-cho, Chikusa-ku, Nagoya 464-8602, Japan
[2]Institute of Liberal Arts and Sciences, Nagoya University, Nagoya 464-8601, Japan

*komori.sachio.h0@f.mail.nagoya-u.ac.jp        *taniyama.tomo@nagoya-u.jp

In a ferromagnet/superconductor/ferromagnet (F/S/F) superconducting spin-valve (SSV), a change in the magnetization alignment of the two F layers modulates the critical temperature ($T_c$) of the S layer. The $T_c$-switching (the SSV effect) is based on the interplay between superconductivity and magnetism. The fast and large resistive switching associated with $T_c$-switching is suitable for nonvolatile cryogenic memory applications. However, the external magnetic field-based operation of SSVs is hindering their miniaturization, and therefore, electric field control of the SSV effect is desired. Here, we report epitaxial growth of a $La_{0.67}Ca_{0.33}MnO_3/YBa_2Cu_3O_7/La_{0.67}Ca_{0.33}MnO_3$ SSV on a piezo-electric $[Pb(Mg_{0.33}Nb_{0.67})O_3]_{0.7}$-$[PbTiO_3]_{0.3}$ (001) substrate and demonstrate electric field control of the SSV effect. Electric field-induced strain-transfer from the piezo-electric substrate increases the magnetization and $T_c$ of the SSV and leads to an enhancement of the magnitude of $T_c$-switching. The results are promising for the development of magnetic-field-free superconducting spintronic devices, in which the S/F interaction is not only sensitive to the magnetization alignment but also to an applied electric field.

## I. INTRODUCTION

The field of spintronics has emerged and rapidly developed following the discovery of the giant magnetoresistance (GMR) effect in a multilayer of a ferromagnet (F) and a nonmagnetic metal (N)[1,2]. Resistive switching of a spin-valve (i.e., a F/N/F trilayer)[3,4] is a basis of the modern spintronic technology and is based on the GMR effect: the electrical resistance increases at the antiparallel (AP) magnetization alignment of the two F layers compared to the parallel (P) alignment due to spin-dependent scattering of spin-polarized electrons. Numerous efforts have been undertaken to decrease the size and energy consumption of spin-valve-based logic and memory devices, and electric field control of ferromagnetism[5,6] and magnetic anisotropy[7–9], which does not require an external magnetic field or electric currents for the resistive switching of spin-valves, is emerging as a promising technology.

Inserting a superconductor (S) instead of N in a spin-valve facilitates a F/S/F superconducting spin-valve (SSV).

A SSV is compatible with cryogenic electronics in which self-heating that changes the properties of superconducting circuits and consumes a limited cooling power of a refrigerator needs to be suppressed. In addition to the normal state resistance of S, the critical temperature ($T_c$) of S is controlled by the magnetization alignment in a SSV[10–12]. Although the mechanism of $T_c$-switching in SSVs depends on material combinations and is complicated, it is broadly categorized into three effects: the magnetic exchange field effect[13–15], the spin scattering effect[16,17] and the stray field effect[18,19]. For a SSV with an *s*-wave S, $T_c$-switching is observed when the S layer thickness ($d_s$) is either comparable to or thinner than the superconducting coherence length ($\xi$)[13–15]. However, in a SSV with a *d*-wave S, $T_c$-switching is observed up to the length scale of $d_s \approx 100\ \xi$[20], which may be due to the nodal superconducting gap with an effectively long $\xi$, enabling $T_c$-switching of a relatively thick *d*-wave S with $T_c$ close to the bulk value. $T_c$-switching of a SSV (the SSV effect) can be used for cryogenic memory devices, which are compatible with energy-efficient superconducting digital circuits and quantum computing circuits[21–24]. However, similar to conventional spin-valves, a technological breakthrough is necessary to realize small devices that do not require external magnetic fields or electric currents.

Here, we report epitaxial growth of a $La_{0.67}Ca_{0.33}MnO_3/YBa_2Cu_3O_7/La_{0.67}Ca_{0.33}MnO_3$ (LCMO/YBCO/LCMO) SSV on a [001]-oriented $[Pb(Mg_{0.33}Nb_{0.67})O_3]_{0.7}\text{-}[PbTiO_3]_{0.3}$ (PMN-PT) substrate and demonstrate a reversible electric field control of the SSV effect. The magnitude of $T_c$-switching is enhanced by up to 6% via electric field-induced strain-transfer from the piezo-electric PMN-PT substrate. The result is promising for the development of electric field controllable superconducting spintronic devices.

## II. EXPERIMEMTS

A LCMO(100 nm)/YBCO(15 nm)/LCMO(50 nm) SSV was epitaxially grown on a commercially available [001]-oriented PMN-PT substrate by pulsed laser deposition (the fourth harmonic of a Q-switched Nd:YAG laser; wavelength $\lambda = 266$ nm). SSVs with LCMO show a large and clear $T_c$-switching with a magnitude up to about 2 K[25–27] and are therefore suitable to investigate an electric field modulation of $T_c$-switching via strain-transfer from PMN-PT with a large piezo-electric constant[28]. Prior to the deposition of the SSV, the PMN-PT substrate was annealed at 633°C for 1 hour and a 20-nm-thick $SrTiO_3$ (STO) buffer layer was grown at the same temperature to prevent the reaction between the substrate and the SSV. The SSV was subsequently grown at 780 °C in 300 mTorr of flowing oxygen. The laser fluence is 0.25 J cm$^{-2}$ for YBCO and 0.5 J cm$^{-2}$ for LCMO and the laser frequency is 10 Hz. After the growth, the SSV was post-annealed at 500°C for 1 h in 600 Torr of oxygen. In-plane electrical resistance ($R$) measurements using a current ($I$) of 100-1000 μA were performed in a Gifford-McMahon cryogen-free system using a four-terminal electrical setup with Au (30 nm)/Ti (5 nm) contacts on the SSVs. A Au/Ti contact was also deposited at the bottom of the substrate to apply an electric field. $R$ was measured as a function of the in-plane magnetic field ($H$), temperature ($T$), and electric field ($E$). $H$ was applied parallel to $I$, and $E = 4$ kV/cm was applied along the [001] direction of PMN-PT at room temperature prior to low temperature measurements. Care was taken to ensure that the leakage current (typically less than 10 nA for a 4.5×4.5 mm$^2$ device area) has no effect on $T_c$ and the resistive switching. We note that $E = 4$ kV/cm is high enough to switch the polarization of PMN-PT[29] but low enough to keep the leakage current below 10 nA. The magnetization ($M$) was measured using a Quantum Design Magnetic Property Measurement System.

## III. RESULTS AND DISCUSSION

We first discuss a magneto-elastic coupling between LCMO and PMN-PT. Figure 1(a) shows a schematic illustration of the polarization directions and strain of PMN-PT with $E$ along the [001] direction. The polarization along the [111], [1$\bar{1}$1], [$\bar{1}$11], and [$\bar{1}\bar{1}$1] directions induced by $E$ leads to a decrease in the $a$- and $b$-axis lattice constants and an increase in the $c$-axis lattice constant. Since the polarization is symmetric with respect to the polarity of $E$, a similar change in the lattice constants is induced for $E$ along the [00$\bar{1}$] direction. Figures 1(b) and 1(c) show in-plane x-ray $2\theta_\chi$–$\varphi$ scan profiles of a LCMO(50 nm)/STO(20 nm)/PMN-PT control sample measured with a grazing-incidence angle of 0.3° around the pseudocubic LCMO (110) and (1$\bar{1}$0) diffraction peaks, respectively. The diffraction peaks shift to a higher angle by applying $E$ = 4 kV/cm, suggesting that an in-plane compressive strain is transferred from the PMN-PT substrate. The lattice spacing of the (110) planes and the (1$\bar{1}$0) planes is calculated to be 2.743 and 2.744 Å, respectively at $E$ = 0, and 2.738 and 2.743 Å, respectively at $E$ = 4 kV/cm. We, therefore, estimate the average in-plane lattice constant to be 3.880 Å at $E$ = 0 and 3.875 Å at $E$ = 4 kV/cm. Figure 1(d) shows out-of-plane x-ray diffraction data from which we estimate the $c$-axis lattice constant of the LCMO layer to be 3.845 Å at $E$ = 0, which is smaller than the in-plane lattice constant, indicating the presence of a lattice mismatch-induced tensile strain along the in-plane. By applying $E$ = 4 kV/cm, the $c$-axis lattice constant of LCMO increases to 3.848 Å, suggesting that the tensile strain along the in-plane is partially relaxed by strain transfer from PMN-PT. The electric field-induced change in the lattice constants (0.11% along the in-plane and 0.072% along the $c$-axis) of the LCMO layer is of the order of that reported for PMN-PT at 4 kV/cm (about 0.05%[29]), implying that the electric field-induced compressive strain along the in-plane is coherently transferred from the PMN-PT substrate to the LCMO layer.

Figure 1(e) shows $M(T)$ curves of the LCMO/STO/PMN-PT sample at $H$ = 5000 Oe (close to the saturation field) for $E$ = 0 and 4 kV/cm. The magnetization of LCMO is increased by 10 emu/cm$^3$ (5.0%) at $E$ = 4 kV/cm compared to $E$ = 0. Similar electric field modulations of the magnetization have been reported for manganite/ferroelectric heterostructures, and several mechanisms have been proposed (e.g., strain-transfer[29–31], carrier doping[29,32], and oxygen migration[33]). We note that an enhancement of the magnetization is observed also at $E$ = − 4 kV/cm in our sample. The symmetric change in the magnetization with respect to the polarity of $E$ rules out the carrier doping effect and the oxygen migration effect, and therefore, the strain-transfer is the most likely origin of the electric field enhancement of the magnetization. Since the hopping integral of $e_g$ electrons responsible for the double exchange interaction in Mn$^{3+}$-O-Mn$^{4+}$ chains is proportional to $\cos^2\varphi/l^{3.5}$, where $\varphi$ is the Mn-O-Mn angle and $l$ is the Mn-O bond length[34], a stronger double exchange interaction is expected under the compressive strain which decreases $l$ without a change in $\varphi$. This is consistent with the electric field enhancement of the magnetization.

We next discuss a LCMO(100 nm)/YBCO(15 nm)/LCMO(50 nm) SSV grown on a PMN-PT (001) substrate. Figure 2(a) shows out-of-plane x-ray diffraction data, which confirm the $c$-axis oriented growth of the SSV and the absence of impurity phases. The $c$-axis lattice constants of LCMO and YBCO are determined to be 3.869Å and 11.55Å, respectively. Rocking curves around the diffraction peaks of the LCMO (002) [Fig. 2(b)] and the YBCO (005) [Fig. 2(c)] show narrow full width at half maximum values of 0.216° and 0.241°, respectively, confirming that the SSV is highly oriented along the [001]-axis of the PMN-PT substrate.

In Fig. 3(a), we plot $R(T)$ of the SSV near the superconducting transition at $E = 0$ and 4 kV/cm, which shows a parallel shift of the $R(T)$ curve indicating an electric field enhancement of $T_c$ down to $R/R_N \approx 10^{-6}$, where $R_N$ is the normal state resistance at the onset temperature of the superconducting transition. An enhancement of $T_c$ via strain-transfer from a (001)-oriented PMN-PT substrate has been reported for YBCO thin films[35]. The $T_c$ enhancement induced by a compressive strain along the in-plane is consistent with a $T_c$ enhancement in YBCO single crystals under uniaxial pressure along the $b$-axis[36]. Figure 3(b) shows $R(H)$ curves at 36 K ($\approx T_c$) for $E = 0$ and 4 kV/cm, where $R$ is normalized at the minimum value. The sharp peaks at $H \approx \pm 200$ Oe indicate a decrease in $T_c$ near the antiparallel magnetization alignment of the two LCMO layers. Since the sign of the resistive switching is opposite to that of the magnetic exchange field effect[16,17] and the stray field effect is negligibly small [see Fig. S3(b) within the Supplemental Material], the switching is likely due to the spin-scattering effect reported for the similar SSVs consisting of YBCO and LCMO[25–27]. The magnitude of the peaks of the normalized $R$ is enhanced by 33% by applying $E = 4$ kV/cm.

To estimate the effective change in $T_c$ resulting from the change in the magnetization alignment ($\Delta T_c$), we compare the magnitude of the resistance peak ($\Delta R$) from the $R(H)$ curve with the slope of the superconducting transition from the $R(T)$ curve [i.e., $\Delta T_c$ is estimated from the relation $\Delta R = \alpha \Delta T_c$, where $\alpha$ is the slope of the $R(T)$ curve at the temperature of the $R(H)$ measurement]. Figure 3(c) shows the temperature dependence of $\Delta T_c$ at $E = 0$ and 4 kV/cm. The $\Delta T_c(T)$ curves show a peak at 31 K for $E = 0$ and 32 K for 4 kV/cm, meaning that the magnitude of the SSV effect is temperature-dependent. The opening of the superconducting gap with decreasing temperature decreases the density of quasiparticles responsible for the spin-scattering while the density of Cooper pairs decreases with increasing temperature. Hence, the quasiparticle-mediated pair breaking effect is maximized at a certain temperature, and this results in the peak feature of the $\Delta T_c(T)$ curves. The electric field-induced shift of the $\Delta T_c(T)$ curve by about 1 K is due to the shift of $T_c$ [shown in Fig. 3(a)]. The maximum $\Delta T_c$ at $E = 4$ kV/cm (700 mK) is higher than that at $E = 0$ (660 mK), meaning that the SSV effect is enhanced by 6% by applying an electric field.

Regardless of the origin of the SSV effect, $\Delta T_c$ can be enhanced by either increasing the magnetization or increasing the maximum magnetization misalignment angle of the two F layers. We note that the coercive fields of the two LCMO layers in our SSVs are comparable. Therefore, the magnetization misalignment angle at $H$ corresponding to the $R$ peak in $R(H)$ is less than 180° and the misalignment angle can be increased if the electric field-induced strain increases the difference of the coercive fields of the two LCMO layers. However, a broadening of the resistive switching in $R(H)$ is not observed at $E = 4$ kV/cm, suggesting that the coercivities are not sensitive to the electric field. Therefore, the likely origin of the enhancement of $\Delta T_c$ is the enhanced magnetization of the two LCMO layers. If this is the case, a similar enhancement of the SSV effect should be observed also in SSVs with other $T_c$-switching mechanisms (e.g., the magnetic exchange effect[13–15,20,37] and the stray field effect[18,19]) and the enhancement can be amplified with decreasing thickness of the S layer, which could be subjects of future investigation.

## IV. CONCLUSION

In conclusion, we have prepared a LCMO/YBCO/LCMO epitaxial SSV on a piezoelectric PMN-PT substrate and demonstrated an electric field enhancement of the SSV effect. Upon application of an electric field, a compressive strain along the in-plane is induced in the SSV and the magnetization of LCMO and $T_c$ of YBCO is enhanced accordingly. These led to an enhanced magnitude of the $T_c$-switching. The electric field control of the S/F interaction demonstrated in this work is a new concept in the field of superconducting spintronics and the results can be potentially applied to various S/F multilayers including magnetic Josephson junctions and are promising for the development of size-scalable superconducting spintronic devices.

## SUPPLEMENTARY MATERIALS

See supplementary materials for the effect of the electric field polarity on the magnetization enhancement of LCMO on PMN-PT(001), and the structural and superconducting properties of YBCO/LCMO on PMN-PT(001).

## ACKNOWLEDGMENTS

The authors acknowledge the funding from JST CREST Grant (No. JPMJCR18J), JSPS KAKENHI Grant (No. JP21H04614, No. JP20K23374, and No. JP23KK0086), and JSPS Bilateral Joint Research Projects Grant (No. JPJSBP120197716). S.K. acknowledges the funding from JST FOREST Grant (No. JPMJFR212V).

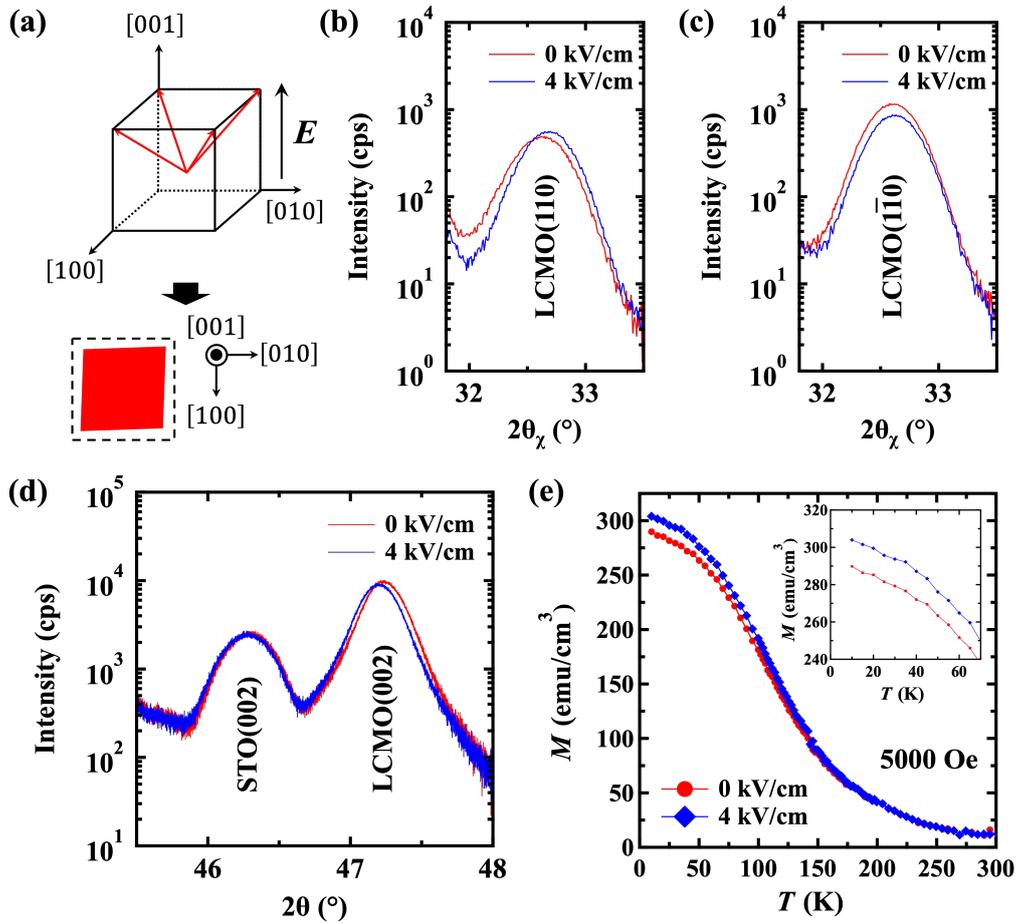

FIG. 1. (a) Schematic illustration of the polarization and lattice strains in PMN-PT with an electric field along the [001] direction. X-ray diffraction patterns around the (b) (110), (c) (1$\bar{1}$0), and (d) (002) pesudocubic peaks of LCMO, and (e) $M(T)$ at $E = 0$ (red curves) and 4 kV/cm (blue curves) for a LCMO(50 nm)/STO(20 nm)/PMN-PT control sample. $M(T)$ was measured during cooling at $H = 5000$ Oe. The inset in (e) shows the magnified $M(T)$ curves.

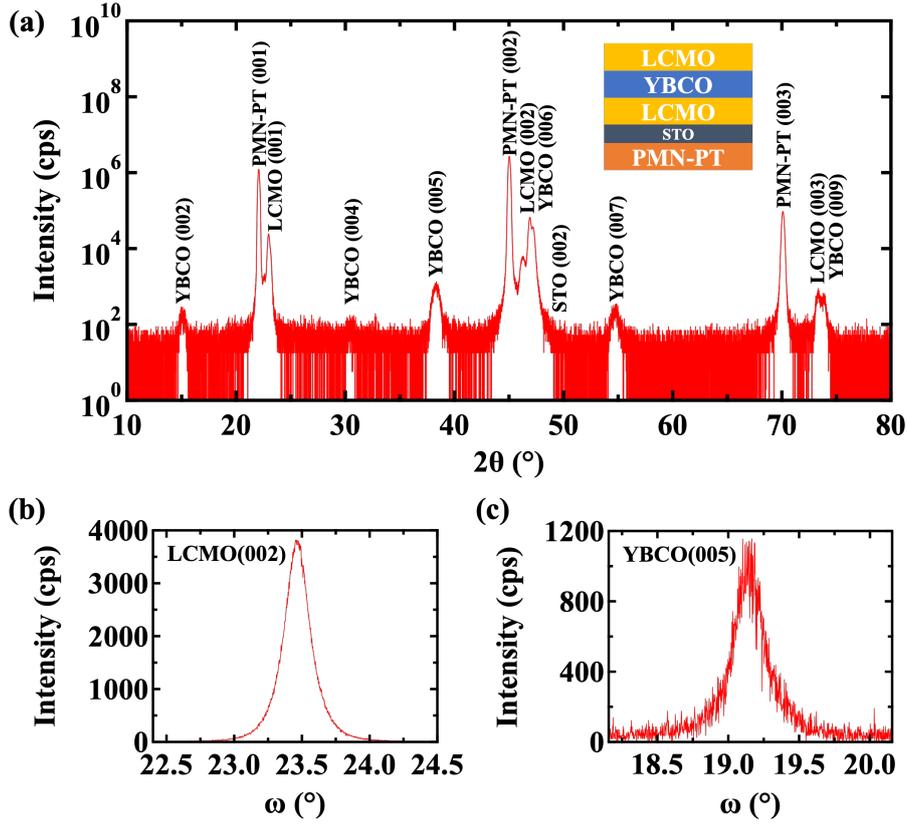

FIG. 2. (a) Out-of-plane x-ray diffraction pattern of LCMO(100 nm)/YBCO(15 nm)/LCMO(50 nm)/STO(20 nm) on a [001]-oriented PMN-PT substrate. Rocking curves on the (b) LCMO (002) and (c) YBCO (005) peaks showing full width at half maximum values of 0.216° and 0.241°, respectively.

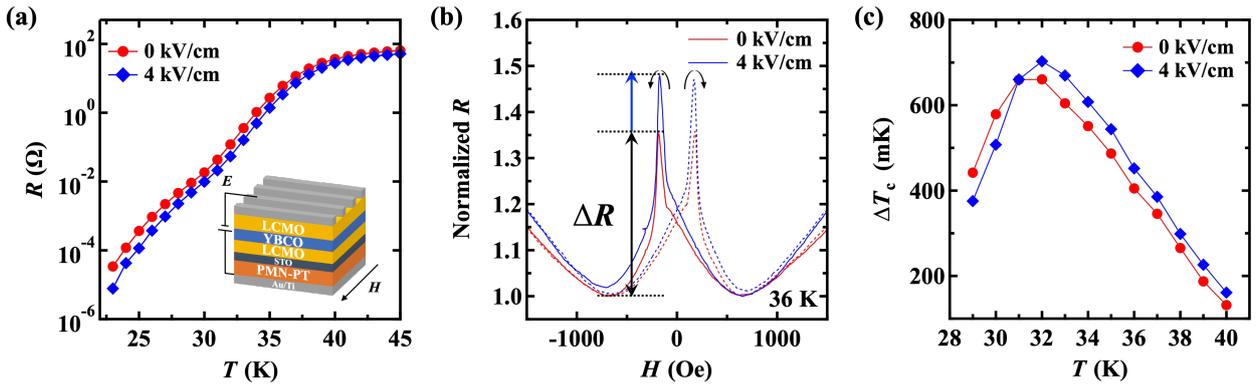

FIG. 3. (a) $R(T)$, (b) $R(H)$, and (c) $\Delta T_c(T)$ curves for LCMO(100 nm)/YBCO(15 nm)/LCMO(50 nm)/STO(20 nm)/PMN-PT at $E = 0$ (red curves) and 4 kV/cm (blue curves). The inset in (a) shows a schematic diagram of the SSV device. The solid and dashed curves in (b) indicate negative and positive $H$-sweeps, respectively.

[1] M.N. Baibich, J.M. Broto, A. Fert, F.N. Van Dau, F. Petroff, P. Eitenne, G. Creuzet, A. Friederich, and J. Chazelas, Phys. Rev. Lett. **61**, 2472 (1988).

[2] G. Binasch, P. Grünberg, F. Saurenbach, and W. Zinn, Phys. Rev. B **39**, 4828 (1989).

[3] B. Dieny, V.S. Speriosu, B.A. Gurney, S.S.P. Parkin, D.R. Wilhoit, K.P. Roche, S. Metin, D.T. Peterson, and S. Nadimi, J. Magn. Magn. Mater. **93**, 101 (1991).

[4] B. Dieny, V.S. Speriosu, S. Metin, S.S.P. Parkin, B.A. Gurney, P. Baumgart, and D.R. Wilhoit, J. Appl. Phys. **4779**, 4774 (1991).

[5] Y.H. Chu, L.W. Martin, M.B. Holcomb, M. Gajek, S.J. Han, Q. He, N. Balke, C.H. Yang, D. Lee, W. Hu, Q. Zhan, P.L. Yang, A. Fraile-Rodríguez, A. Scholl, S.X. Wang, and R. Ramesh, Nat. Mater. **7**, 478 (2008).

[6] S.M. Wu, S. Cybart, P. Yu, M.D. Rossell, J.X. Zhang, R. Ramesh, and R.C. Dynes, Nat. Mater. **9**, 756 (2010).

[7] T. Maruyama, Y. Shiota, T. Nozaki, K. Ohta, N. Toda, M. Mizuguchi, A.A. Tulapurkar, T. Shinjo, M. Shiraishi, S. Mizukami, Y. Ando, and Y. Suzuki, Nat. Nanotechnol. **4**, 158 (2009).

[8] T. Nozaki, T. Yamamoto, S. Miwa, and M. Tsujikawa, Micromachines **10**, 327 (2019).

[9] T. Taniyama, J. Phys. Condens. Matter **27**, 504001 (2015).

[10] S. Oh, D. Youm, and M.R. Beasley, Appl. Phys. Lett. **71**, 2376 (1997).

[11] L.R. Tagirov, Phys. Rev. Lett. **83**, 2058 (1999).

[12] A.I. Buzdin, Vedyayev, A, V, and Ryzhanova, N, V, Europhys. Lett. **48**, 686 (1999).

[13] P.G. De Gennes, Phys. Lett. **23**, 10 (1966).

[14] B. Li, N. Roschewsky, B.A. Assaf, M. Eich, M. Epstein-Martin, D. Heiman, M. Münzenberg, and J.S. Moodera, Phys. Rev. Lett. **110**, 097001 (2013).

[15] Y. Zhu, A. Pal, M.G. Blamire, and Z.H. Barber, Nat. Mater. **16**, 195 (2016).

[16] A. Singh, C. Sürgers, R. Hoffmann, H. V. Löhneysen, T. V. Ashworth, N. Pilet, and H.J. Hug, Appl. Phys. Lett. **91**, 152504 (2007).

[17] A.Y. Rusanov, S. Habraken, and J. Aarts, Phys. Rev. B **73**, 060505(R) (2006).

[18] J. Zhu, X. Cheng, C. Boone, and I.N. Krivorotov, Phys. Rev. Lett. **103**, 027004 (2009).

[19] D. Stamopoulos, E. Manios, and M. Pissas, Phys. Rev. B **75**, 184504 (2007).

[20] A. Di Bernardo, S. Komori, G. Livanas, G. Divitini, P. Gentile, M. Cuoco, and J.W.A. Robinson, Nat. Mater. **18**, 1194 (2019).

[21] I.M. Dayton, T. Sage, E.C. Gingrich, M.G. Loving, T.F. Ambrose, N.P. Siwak, S. Keebaugh, C. Kirby, D.L. Miller, A.Y. Herr, Q.P. Herr, and O. Naaman, IEEE Magn. Lett. **9**, 3301905 (2018).

[22] I.I. Soloviev, N. V Klenov, S. V Bakurskiy, M.Y. Kupriyanov, A.L. Gudkov, and A.S. Sidorenko, Beilstein J. Nanotechnol. **8**, 2689 (2017).

[23] Y. He, J. Li, Q. Wang, H. Matsuki, and G. Yang, Adv. Devices Instrum. **4**, 0035 (2023).

[24] J. Linder and J.W.A. Robinson, Nat. Phys. **11**, 307 (2015).

[25] V. Peña, Z. Sefrioui, D. Arias, C. Leon, J. Santamaria, J.L. Martinez, S.G.E. Te Velthuis, and A. Hoffmann, Phys. Rev. Lett. **94**, 057002 (2005).

[26] N.M. Nemes, M. García-Hernández, S.G.E. Te Velthuis, A. Hoffmann, C. Visani, J. Garcia-Barriocanal, V. Peña, D. Arias, Z. Sefrioui, C. Leon, and J. Santamaría, Phys. Rev. B **78**, 094515 (2008).

[27] N.M. Nemes, C. Visani, C. Leon, M. Garcia-Hernandez, F. Simon, T. Fehér, S.G.E. te Velthuis, A. Hoffmann, and J. Santamaria, Appl. Phys. Lett. **97**, 032501 (2010).

[28] M. Shanthi, L.C. Lim, K.K. Rajan, and J. Jin, Appl. Phys. Lett. **92**, 142906 (2008).

[29] Z.G. Sheng, J. Gao, and Y.P. Sun, Phys. Rev. B **79**, 174437 (2009).

[30] W. Eerenstein, M. Wiora, J.L. Prieto, J.F. Scott, and N.D. Mathur, Nat. Mater. **6**, 348 (2007).

[31] C. Thiele, K. Dörr, O. Bilani, J. Rödel, and L. Schultz, Phys. Rev. B **75**, 054408 (2007).

[32] H. Lu, T.A. George, Y. Wang, I. Ketsman, J.D. Burton, S. Ryu, D.J. Kim, J. Wang, C. Binek, P.A. Dowben, A. Sokolov, E.Y. Tsymbal, and A. Gruverman, Appl. Phys. Lett. **100**, 232904 (2012).

[33] K. Imura, S. Ishikawa, S. Komori, and T. Taniyama, Appl. Phys. Lett. **122**, 202402 (2023).

[34] D.P. Kozlenko and B.N. Savenko, Phys. Part. Nucl. **37**, S1 (2006).

[35] P. Pahlke, S. Trommler, B. Holzapfel, L. Schultz, and R. Hühne, J. Appl. Phys. **113**, 123907 (2013).

[36] W.H. Fietz, K.P. Weiss, and S.I. Schlachter, Supercond. Sci. Technol. **18**, S332 (2005).

[37] S. Komori, A. Di Bernardo, A.I. Buzdin, M.G. Blamire, and J.W.A. Robinson, Phys. Rev. Lett. **121**, 077003 (2018).

# Supplementary Information for

# Electric field enhancement of the superconducting spin-valve effect via strain-transfer across a ferromagnetic/ferroelectric interface

Tomohiro Kikuta[1], Sachio Komori[1*], Keiichiro Imura[1,2], Tomoyasu Taniyama[1*]

[1]Department of Physics, Nagoya University, Furo-cho, Chikusa-ku, Nagoya 464-8602, Japan
[2]Institute of Liberal Arts and Sciences, Nagoya University, Nagoya 464-8601, Japan

*komori.sachio.h0@f.mail.nagoya-u.ac.jp          *taniyama.tomo@nagoya-u.jp

## 1. Effect of the electric field polarity on the magnetization enhancement of LCMO on PMN-PT (001)

In Fig. S1, we plot $M(H)$ curves for LCMO(50 nm)/STO(20 nm)/PMN-PT at $E = 0$, $+4$, and $-4$ kV/cm. The manetization at 20 K is enhanced by 5.0% at $E = +4$ kV/cm and 4.4% at $E = -4$ kV/cm compared with the value at $E = 0$. The symmetric enhancement of the magnetization with respect to the polarity of $E$ rules out the oxygen migration and carrier doping effects as its origin.

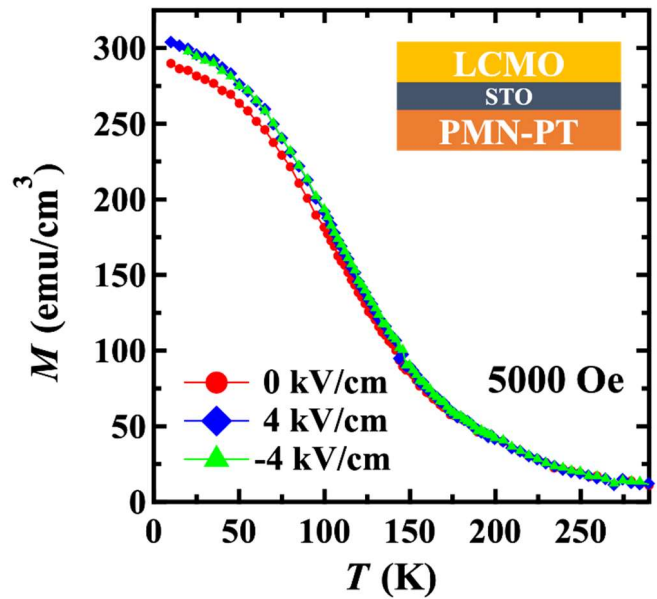

FIG. S1. Temperature dependence of the magnetization for LCMO(50 nm)/STO(20 nm)/PMN-PT at $H = 5000$ Oe for $E = 0$ (red curve), $+4$ (blue curve), and $-4$ kV/cm (green curve).

## 2. X-ray diffraction pattern of a YBCO/LCMO bilayer on PMN-PT(001)

Figure S2 shows out-of-plane x-ray diffraction data from STO(5 nm)/YBCO(15 nm)/LCMO(50 nm)/STO(20 nm)/PMN-PT, confirming $c$-axis oriented growth of YBCO and LCMO. We note that the 5-nm-thick STO capping layer was deposited to prevent out-diffusion of oxygen from YBCO and the 20-nm-thick STO buffer layer was deposited to prevent the reaction between LCMO and PMN-PT.

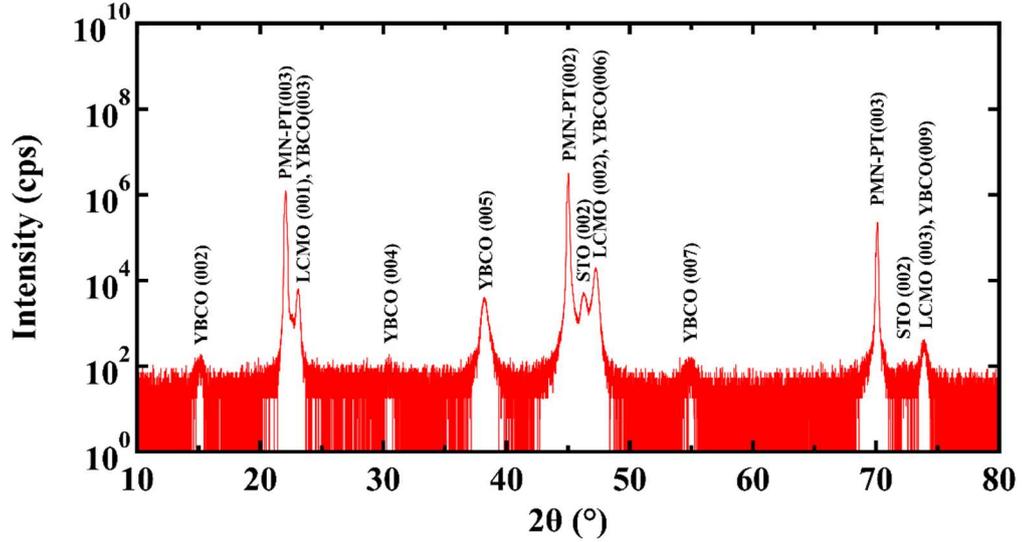

FIG. S2. Out-of-plane x-ray diffraction pattern of a YBCO(15 nm)/LCMO(50 nm) bilayer on a PMN-PT(001) substrate.

## 3. Superconducting properties of a YBCO/LCMO bilayer on PMN-PT(001)

Figure S3(a) shows $R(T)$ curves of the YBCO/LCMO bilayer at $E = 0$ and 4 kV/cm. $T_c$ of the bilayer is higher than that of the superconducting spin-valve in the main manuscript. This is probably due to the absence of the growth process of the 100-nm-thick top LCMO layer causing out-diffusion of oxygen from YBCO. Since $T_c$ of the bilayer is close to the optimum value, the electric field enhancement of $T_c$ is smaller than that observed for the superconducting spin-valve in the main manuscript. Figure S3(b) shows $R(H)$ curves of the bilayer at 72 K ($\approx T_c$) for $E = 0$ and 4 kV/cm. The monotonic increase of $R$ with $H$ is due to the field suppression of the superconductivity. The absence of the resistive switching near the coercive field ($H \approx \pm 200$ Oe) suggests that the stray field effect from magnetic domain walls of LCMO is negligibly small.

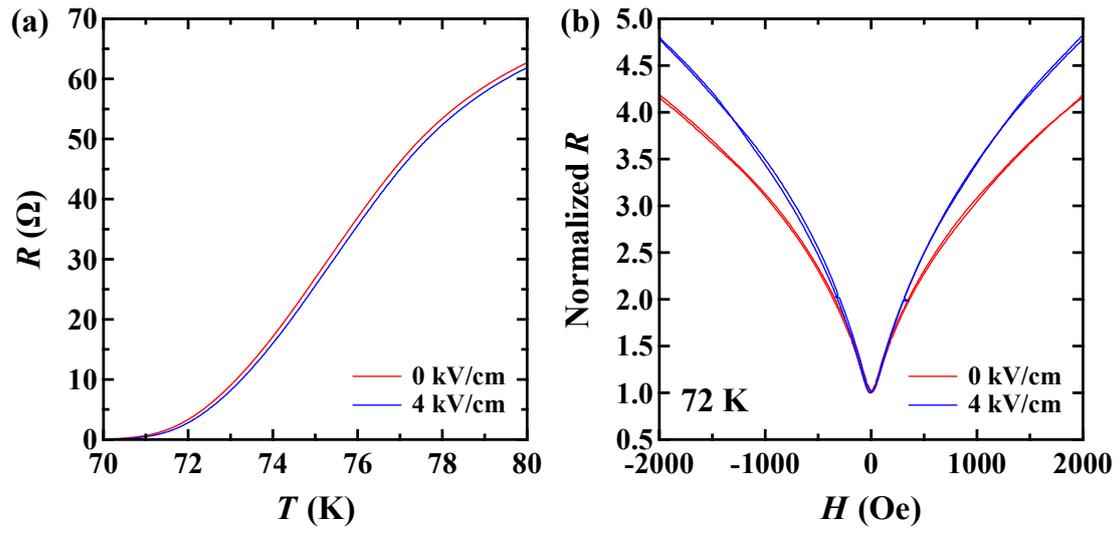

FIG. S3. (a) $R(T)$ and (b) $R(H)$ at 72 K for a YBCO(15 nm)/LCMO(50 nm) bilayer on PMN-PT(001) at $E = 0$ (red curves) and 4 kV/cm (blue curves).